\documentclass[preprint,12pt]{elsarticle}

\usepackage{amssymb}  
\usepackage{amsmath}  
\usepackage{graphicx} 
\usepackage{booktabs} 
\usepackage{geometry} 
\usepackage{lineno}   
\usepackage{natbib}
\usepackage{multirow}
\usepackage{xcolor}
\usepackage[table]{xcolor}  
\usepackage{multirow, booktabs} 
\usepackage{url}
\usepackage{pifont}
\newcommand{\cmark}{\ding{51}}%
\newcommand{\xmark}{\ding{55}}%
\journal{Computer Speech \& Language}

\begin{document}

\begin{frontmatter}

\title{Unmasking Deepfakes: Leveraging Augmentations and Features Variability for Deepfake Speech Detection}

\author[inst1]{Inbal Rimon\corref{cor1}}
\ead{inbalri@post.bgu.ac.il}
\author[inst2]{Oren Gal}
\author[inst1]{Haim Permuter}

\cortext[cor1]{Corresponding author}

\address[inst1]{Ben Gurion University, Be'er Sheva, Israel}
\address[inst2]{University of Haifa, Haifa, Israel}

\begin{abstract}
Deepfake speech detection presents a growing challenge as generative audio technologies continue to advance. 
We propose a hybrid training framework that advances detection performance through novel augmentation strategies. 
First, we introduce a dual-stage masking approach that operates both at the spectrogram level (MaskedSpec) and within the latent feature space (MaskedFeature), providing complementary regularization that improves tolerance to localized distortions and enhances generalization learning. 
Second, we introduce compression-aware strategy during self-supervised to increase variability in low-resource scenarios while preserving the integrity of learned representations, thereby improving the suitability of pretrained features for deepfake detection. 
The framework integrates a learnable self-supervised feature extractor with a ResNet classification head in a unified training pipeline, enabling joint adaptation of acoustic representations and discriminative patterns. 
On the ASVspoof5 Challenge (Track~1), the system achieves state-of-the-art results with an Equal Error Rate (EER) of 4.08\% under closed conditions, further reduced to 2.71\% through fusion of models with diverse pretrained feature extractors. 
when trained on ASVspoof2019, our system obtaining leading performance on the ASVspoof2019 evaluation set (0.18\% EER) and the ASVspoof2021 DF task (2.92\% EER). 
\end{abstract}

\begin{keyword}
Deepfake Speech Detection \sep Speech Processing \sep ASVSpoof5 \sep Speech Augmentations \sep Speech Features 
\end{keyword}

\end{frontmatter}

\section{Introduction}

\subsection{Deepfake Speech Detection}
Deepfake speech poses serious security concerns across various fields, including cybersecurity, law enforcement, and military operations. Synthetic audio can be exploited for misinformation, impersonation, and fraud, necessitating detection techniques to ensure the integrity of audio-based communications. Early detection systems relied on handcrafted features such as Mel-frequency cepstral coefficients (MFCCs) and spectrograms, typically processed by classifiers like support vector machines (SVMs) or Gaussian mixture models (GMMs) \cite{mcuba2023effect, Sahidullah2015, yi2023audio, alzantot2019deep}. These approaches offered moderate success but lacked robustness against evolving generative deepfake techniques. More recent methods have adopted convolutional neural networks (CNNs), which can effectively capture local time-frequency patterns and subtle artifacts left by generative models. Architectures such as ResNet \cite{alzantot2019deep, Rimon2022}  and SENet \cite{tak2022automatic} have shown strong performance on benchmark datasets, benefiting from their ability to model fine-grained acoustic distortions.

The ASVSpoof challenge series \cite{ASVSpoof2019, ASVSpoof2021} has become a critical benchmark for evaluating deepfake detection systems. The latest edition, ASVSpoof5 (ASVSpoof2024), includes two tracks: 
Track 1 focuses on standalone deepfake speech detection, independent of ASV systems, while Track 2 integrates detection within ASV pipelines \cite{ASVSpoof5}. ASVSpoof5 introduces diverse spoofing techniques, codecs, and speaker variations to reflect practical deployment scenarios. Performance is typically evaluated using metrics such as Equal Error Rate (EER), encouraging the development of detection systems that generalize across unseen attack types.

\subsection{Speech Features}
Feature extraction remains critical to effective deepfake detection. Traditional representations such as spectrograms offer a time-frequency view of the signal \cite{oppenheim1999discrete} and are widely used as CNN inputs. While these features have been instrumental in early systems \cite{mittal2022automatic, li2024audio}, the use of spectrograms may be approaching a performance ceiling. Recent advances suggest that further improvements in deepfake detection may require more sophisticated feature extraction methods, as the complexity of spoofing attacks continues to increase.

Recent studies have explored the use of pre-trained models for speech representation learning, particularly Wav2Vec 2.0 \cite{baevski2020wav2vec} and WavLM \cite{chen2022wavlm}. These models were originally developed for tasks such as automatic speech recognition (ASR) and speaker verification, and are trained on large-scale unlabeled speech corpora using masked prediction objectives. While not explicitly designed for deepfake detection, their learned representations have been shown to be effective in this context, as they capture rich acoustic and speaker-discriminative features that can help identify subtle artifacts introduced by synthetic speech generation \cite{tak2022automatic, xie2021siamese}.

\subsection{Speech Augmentations}

Data augmentation techniques play a pivotal role in stability and broad applicability of deepfake speech detection models. By introducing variations to the training data, augmentation methods aim to simulate real-world conditions, enabling the models to learn generalized patterns that can better handle unseen, adversarial, or out-of-distribution inputs. Common speech augmentation techniques include pitch shifting, time-stretching, noise addition, and speed perturbation. These methods help models adapt to different speaker characteristics, environmental conditions, and recording qualities \cite{yi2023audio, zeng2023deepfake}.

Beyond speech augmenting, additional methods focus on audio augmenting. SpecAugment \cite{park2019specaugment}, developed for speech recognition, enhances generalization by masking random time and frequency blocks in the spectrogram. A further adaptation, SpecAverage \cite{Rimon2022}, was proposed to address the non-zero mean characteristics of audio features, making it more suitable for deepfake speech detection. However, a limitation of both SpecAugment and SpecAverage is that the applied masks have a fixed shape, which may not adequately represent the complex and dynamic noise distributions encountered in real-world speech data. This constraint arises from the fact that the mask shapes are primarily designed for computational efficiency rather than mimic audio distortion patterns.

Another general-purpose augmentation technique is RawBoost \cite{tak2022automatic}, which applies perturbations directly in the waveform domain, such as amplitude scaling, reverberation, and filtering. Additionally, low-pass filtering (LPF) has been shown to enhance detection performance by attenuating high-frequency components, which often contain residual artifacts introduced during speech synthesis.

\subsection{Main Contributions}
This work presents a deepfake speech detection model trained end-to-end for the ASVSpoof5 challenge, achieving state-of-the-art performance in Track 1 under the closed condition setting. The study focuses on augmentation and training strategies designed to enhance the model’s ability to perform well despite the limited size and linguistic scope of the training data. The primary contributions of this work are summarized below:

\begin{enumerate}
    \item \textbf{Two-stage masking framework:} We design an augmentation strategy that combines spectral masking (MaskedSpec) and latent-space masking (MaskedFeature). Multiple mask types are evaluated, and the best-performing configurations consistently improve robustness and provide additional gains when combined with established methods such as RawBoost and compression augmentation.
    
\item \textbf{Compression-aware self-supervised pretraining:} Introduces lossy codec perturbations directly into the self-supervised pretraining stage, unlike prior work that applies them only during supervised fine-tuning. This strategy strengthens the resilience of self-supervised representation learning for deepfake detection, particularly when training data is limited, by increasing data diversity without disrupting the underlying audio structure.

\item \textbf{Hybrid training framework:} We present a novel integration of a self-supervised, learnable feature extractor with a convolutional classifier (ResNet34). While each component is well established individually, our framework couples them within a unified hybrid training process and optimizes them jointly in an end-to-end scheme. This design enables task-specific adaptation of both low-level acoustic representations and high-level discriminative features.

\end{enumerate}

\section{Method}

This section outlines the methodology we employed, beginning with our proposed hybrid pipeline, which combines a trainable feature extraction module (Wav2Vec2) with a classification head (ResNet34). Our approach incorporates augmentation strategies applied at different stages of the pipeline to enhance model performance. For raw audio, we employ \textit{MaskedSpec}, low-pass filtering (LPF), rawboost and compression augmentation. Compression augmentations, in particular, are utilized both during the pretraining phase of the feature extractor and in the end-to-end training process. At the feature level, our work introduces \textit{MaskedFeature} and feature normalization.

\begin{figure}[h!]
\centering
\includegraphics[width=0.8\textwidth]{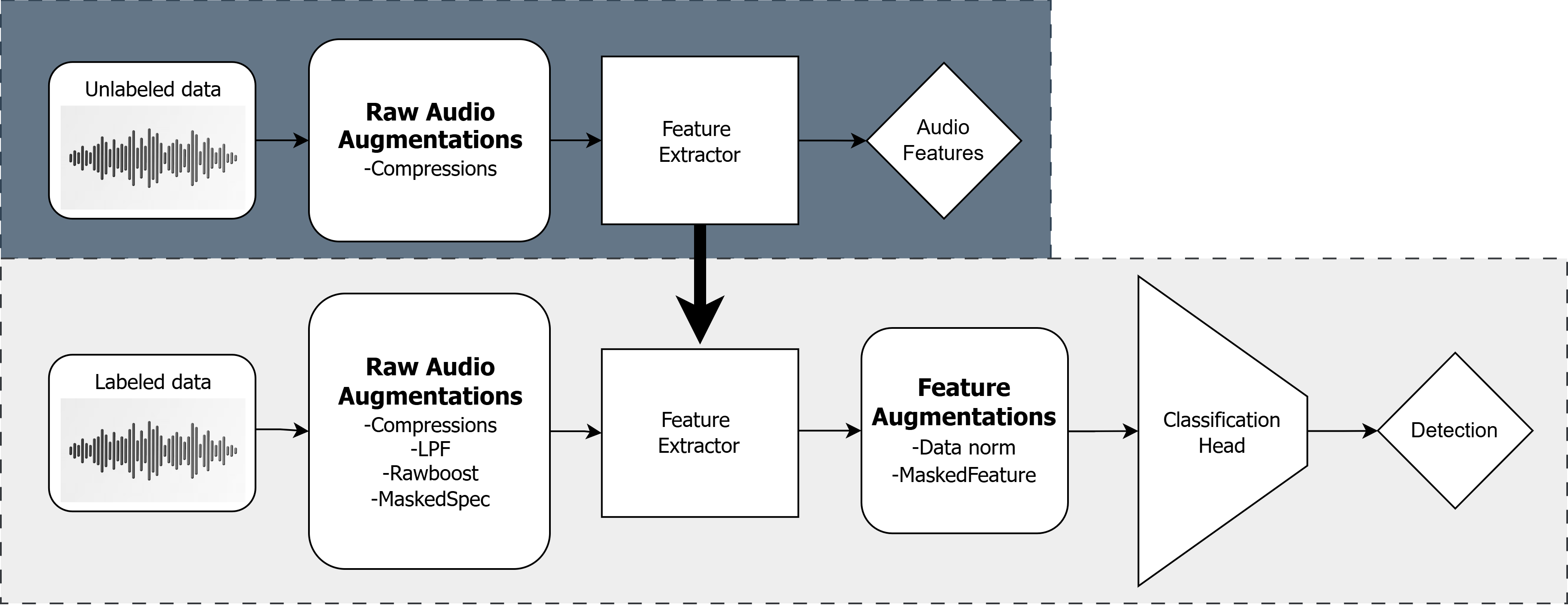} 
\caption{\it Overview of the proposed hybrid training framework. In the first stage (top, dark-shaded blocks), raw audio is used to pretrain the feature extractor using a self-supervised objective. In the second stage (bottom, light-shaded blocks), the model is trained end-to-end for the task of deepfake speech detection. Augmentations are applied at multiple points in the pipeline to boost tolerance to variability and to enrich the training signal under low-resource condition. The feature extractor is initialized with weights from the pretraining phase, as indicated by the downward arrow.}
\label{model}
\end{figure}

\subsection{Hybrid Training Pipeline}

We propose a hybrid training framework for deepfake speech detection that integrates self-supervised pretraining with supervised end-to-end training. As illustrated in Figure~\ref{model}, the system consists of a trainable feature extractor and a classification head, with augmentations applied during both stages. In the first phase, the feature extractor is pretrained on unlabeled audio using a self-supervised objective. To enhance robustness in low-resource conditions, we introduce compression-based augmentations at this stage, promoting invariance in learned representations and improving feature transferability across domains.
In the supervised stage, the pretrained feature extractor is initialized with the learned weights and jointly fine-tuned with the classification head using labeled data. This stage incorporates additional raw audio augmentations, including low-pass filtering, RawBoost perturbations, and spectrogram masking, as well as feature-level augmentations such as normalization and feature masking. These augmentations improve generalization by exposing the model to signal variations encountered in real-world audio. We use Wav2Vec 2.0 as the feature extractor and ResNet34 as the classification head, leveraging their respective strengths in acoustic representation and deep pattern modeling. The end-to-end training strategy enables the model to adapt both low-level and high-level representations to the task of detecting synthetic speech.
An overview of the full pipeline is shown in Figure~\ref{model}.

\subsection{Augmentation Strategies}

To enhance generalization to unseen deepfake techniques and codec conditions, we employ a comprehensive augmentation strategy applied at different stages of training \footnote{\url{https://github.com/InbalRim/MaskedSpec}}.  

During supervised training, we introduce a two-stage masking framework that operates at both the waveform and feature levels. These masking strategies are conceptually grounded in dropout theory, where masking acts as structured regularization. At the spectral level, masks occlude frequency regions, forcing the model to reconstruct missing information from context. At the feature level, masking suppresses localized latent activations, encouraging distributed and noise-tolerant embeddings.  

In the self-supervised stage, we propose compression-aware learning by injecting lossy codec artifacts directly into the input. 
This approach incorporates codec variability directly into the representation learning phase, serving as a natural form of data augmentation that expands the effective training set without introducing artificial distortions that might mislead the feature extractor. In low-resource, single-language scenarios, this strategy provides a principled way to improve resilience for deepfake detection while preserving the integrity of the underlying speech signal.

Beyond these core contributions, we also evaluate a range of additional augmentations to assess their individual and combined effects. Together, these methods form a layered augmentation framework that promotes robustness to diverse perturbations and distributional shifts.

\subsubsection{Raw Audio Augmentations}

\begin{figure}[h!]
\centering
\includegraphics[width=0.5\textwidth]{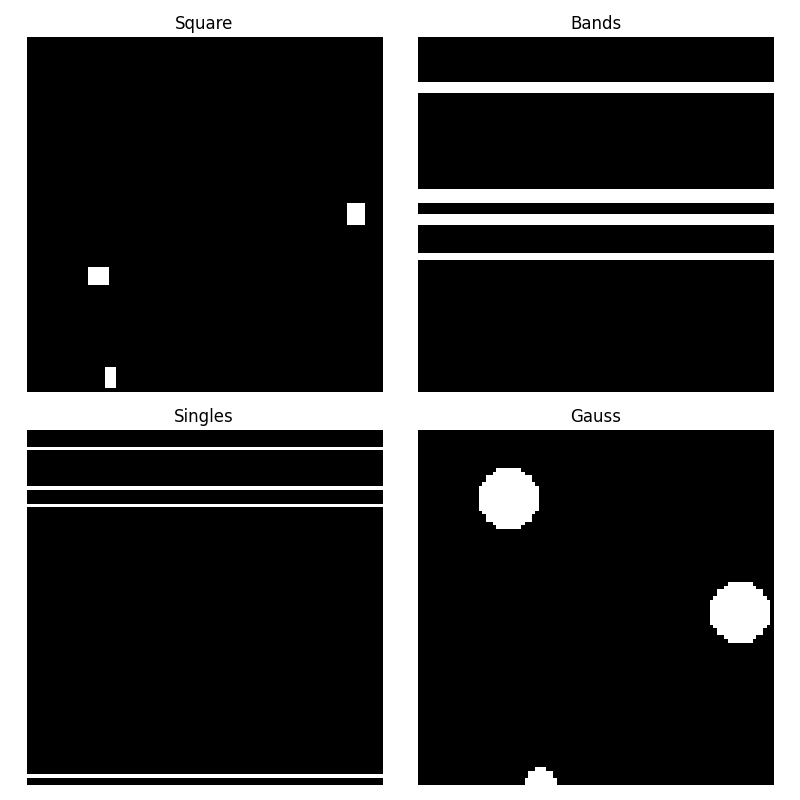} 
\caption{\it Visualization of the different mask types used in our experiments: Squares, Bands, Singles, and Gauss. White regions indicate the masking value ($\mu_{\text{stft}}$), while black regions represent the original, unmasked values.}
\label{mask_types}
\end{figure}

Inspired by SpecAugment \cite{park2019specaugment} and SpecAverage \cite{Rimon2022}, we sought to enhance the training dataset by performing augmentations on the raw audio through its frequency domain representation. 
The proposed \textbf{MaskedSpec} technique employs frequency and time-domain masking to the audio's Short-Time Fourier Transform (STFT). First, the STFT of the raw audio signal is calculated using the formula:  
\[
X(m, k) = \sum_{n=0}^{N-1} x(n+m) \cdot w(n) \cdot e^{-j2\pi kn/N},
\]
where \( x(n) \) is the input audio, \( w(n-m) \) represents the window function, \( m \) and \( k \) are the time and frequency bins, and \( N \) is the frame size.  

Next, the mean value of the STFT is computed. Initially, the mean magnitude and mean phase of the STFT are calculated as follows:

\begin{equation}
\mu_{\text{magnitude}} = \frac{1}{N} \sum_{n=1}^{N} |x(n)|
\end{equation}

\begin{equation}
\mu_{\text{phase}} = \frac{1}{N} \sum_{n=1}^{N} \arg(x(n))
\end{equation}
Then, a complex value is reconstructed from these means:

\begin{equation}
\mu_{\text{stft}} = \mu_{\text{magnitude}}  \cdot \exp(j \cdot \mu_{\text{phase}})
\end{equation}
\(\mu_{\text{stft}}\) is a complex number that serves as the \textit{masking value} to occlude portions of the STFT. After the masks are applied, the inverse Short-Time Fourier Transform (iSTFT) is used to convert the modified STFT back into the time domain, and the resulting signal is then passed as input into the feature extractor for further processing.

Several masking shapes were explored to optimize the model's capacity for generalization: \textit{Square} masks occlude localized regions of the spectrogram, encouraging reliance on global patterns over specific, localized features.
\textit{Bands} masks introduce elongated occlusions across continuous frequency ranges, mimicking spectral loss and requiring the model to capture dependencies over extended temporal or spectral contexts.
\textit{Singles} masks target individual frequency bands, providing simpler occlusions that still effectively challenge the model. \textit{Gaussian} mask gradually degrade the spectrogram, simulating natural signal distortions such as noise-induced corruption or environmental degradation. While traditional masking strategies are often favored for their computational efficiency and ease of implementation, we sought to explore masks that, although more computationally demanding, could offer improved relevance and effectiveness for the task.
These varied masking strategies aim to expose the model to a wide range of distortions, improving its robustness to unseen conditions. A visualization of these masking types is presented in Figure \ref{mask_types}.

\textbf{Low-pass filtering (LPF)} was used as a lightweight augmentation to simulate high-frequency loss commonly introduced by codecs or suboptimal recording conditions. By randomly varying the cutoff frequency, this technique injects spectral variability, helping the model generalize to audio degraded in different ways. LPF was implemented via 1D convolution with a Hamming-windowed sinc kernel:
\begin{equation}
h[m] = \frac{\sin(2\pi f_{\text{cutoff}} m)}{\pi m} \cdot \text{window}[m], \quad h[0] = 2 \cdot f_{\text{cutoff}}.
\end{equation}

To increase the diversity of distortion types during training, we incorporate \textbf{RawBoost} as an additional waveform-level augmentation. RawBoost introduces a sequence of audio perturbations, including amplitude scaling, filtering, and reverberation, to simulate realistic acoustic degradations commonly found in practical scenarios \cite{tak2022automatic}. We adopt the configuration recommended by the original authors, using the prescribed combination of distortion functions.

To improve robustness against real-world audio distortions, we incorporate an additional \textbf{compression-decompression} (codec) augmentation into both the self-supervised pretraining and supervised end-to-end training phases. Lossy codecs such as MP3 and M4A, commonly used for storage and streaming, introduce artifacts including quantization noise, spectral smearing, and high-frequency attenuation. These distortions can hinder detection performance, especially in mismatched conditions.
Our augmentation simulates such effects by encoding and decoding audio at varying bitrates, enriching the dataset with realistic degradation. In the supervised stage, this exposes the model to codec-induced variability, helping it learn discriminative features that are robust to compression artifacts. During self-supervised pretraining, this additional variation improves representation learning by broadening the acoustic diversity encountered by the feature extractor.
To maintain computational feasibility under ASVSpoof5 challenge constraints, we selected a limited yet representative set of compression formats and bitrates. This design choice was guided by prior work \cite{Rimon2022}, which demonstrated that training with a small but diverse set of compressions can generalize well to unseen codec conditions without incurring prohibitive computational cost.

\subsubsection{Feature Augmentations}

To further expose the model to potential feature artifacts and improve its generalization ability, we employed an augmentation strategy similar to \textit{MaskedSpec}, but applied directly to the latent representations produced by the feature extractor. This augmentation is referred to as \textbf{MaskedFeature}. The primary goal of this technique is to force the model to rely on broader, more generalized patterns in the data, rather than focusing on specific, localized details that might be overly sensitive to noise or distortions.
In \textit{MaskedFeature}, portions of the feature space are occluded during training, which simulates the presence of distortions or information loss in the latent representations. By masking these parts of the feature space, the model learns to adapt to a more holistic view of the input data, promoting robustness and resilience. This approach is beneficial in scenarios where certain features of the input may be corrupted or unavailable, such as in deepfake detection tasks where some parts of the signal may be intentionally manipulated. We also investigate the use of various mask shapes for \textbf{MaskedFeature}, aiming to further explore how different patterns of occlusion influence the model’s ability to generalize.

\textbf{Feature normalization} was implemented to scale the extracted feature values into a consistent range, such as $[-1, 1]$. Normalization minimizes the dominance of individual features, promoting balanced learning and reducing sensitivity to recording inconsistencies or device-specific biases. This step is critical for ensuring robustness across varied input conditions.

\section{Results and Insights}
This section presents a detailed evaluation of the proposed deepfake detection framework, focusing on the contribution of individual components and the impact of the proposed augmentation strategies. We begin by assessing the effect of various masking configurations at both the spectrogram and feature levels. We then evaluate the role of compression and RawBoost augmentations under different training regimes.
Fusion analyses and ablation studies are included to further understand the individual and joint effects of the system components.
Performance is reported primarily in terms of Equal Error Rate (EER), additional results, including evaluations using minimum Detection Cost Function (minDCF), are provided in the Appendix for completeness.

\subsection{Experiments Setup}
Experiments were conducted using the proposed architecture model, exploring the effects of various data augmentation and manipulation techniques. Models trained on a GPU with 24 GB of memory. The training utilized a constant batch size of 16, and the number of epochs was limited to 5 to prevent overfitting while addressing time constraints. Compression augmentations employed MP3 with a bitrate of 16 kbps and M4A at 64 kbps during training. For development, we used MP3 at 48 kbps along with M4A at both 16 kbps and 64 kbps. Although development augmentations are typically not performed, their inclusion in this case enhances the assessment of the model's robustness. LPF was applied online, with a randomly selected frequency cutoff. For both MaskedSpec and MaskedFeature, masks were generated online, with a random selection of the number and size of each patch. Although creating masks online is time-consuming, this approach is valuable for capturing diverse aspects of data manipulation.Pretrained feature extractors were used solely for initialization, with the entire model, including the feature extractor, subsequently trained end-to-end throughout the training process. The evaluation is conducted using the ASVSpoof5 datasets, where the Equal Error Rate (EER) for both the development (Dev EER) and evaluation (Eval EER) sets is calculated. Although other metrics within the ASVSpoof5 framework provide useful insights, we prioritize EER due to its significance in assessing the model's ability to accurately differentiate between genuine and deepfake speech.

\subsection{Model Architecture and Features}
Trainable SSL features substantially outperform traditional spectrogram inputs, and self-supervised pretraining provides a significant boost in both development and evaluation EER.

\begin{table}[h!]
\centering
\begin{tabular}{@{}ccc@{}}
\toprule
Features & Dev EER (\%) & Eval EER (\%) \\
\midrule
Spectrogram &  47.86 & 38.72 \\
Wav2Vec2 (without pretraining) &  38.24 & 40.64  \\
Wav2Vec2 (pretrained on ASVSpoof5) & \textbf{20.82} & \textbf{26.36} \\

\bottomrule
\end{tabular}
\caption{Performance of classification head with different features.}
\label{features_results}
\end{table}

We explore the performance differences between traditional, non-trainable features such as spectrograms and trainable feature extractors like Wav2Vec2, using the same classification head, ResNet34. Traditional features like spectrograms provide a static representation of audio based on predefined transformations, requiring the classifier to identify meaningful patterns from these fixed inputs. On the other hand, trainable feature extractors, such as Wav2Vec2, dynamically learn task-specific representations directly from raw audio, leveraging deep learning's ability to extract hierarchical features. 

The results in Table \ref{features_results} demonstrate the clear advantages of trainable features over traditional spectrogram-based representations. Models trained with spectrogram features exhibit the weakest performance, with a Dev EER of 47.86\% and Eval EER of 38.72\%. This indicates that spectrograms, without additional learned enhancements, struggle to effectively separate genuine from fake audio in this task. Introducing Wav2Vec2 as the feature extractor significantly reduces the EER, even without pretraining, achieving a Dev EER of 38.24\% and Eval EER of 40.64\%. Despite this improvement, the lack of domain-specific pretraining likely limits its generalization capabilities.

Significant improvement is observed when the model is trained in two distinct phases. Initially, Wav2Vec2 undergoes self-supervised pertaining, as detailed in \cite{fairseq}, on the ASVSpoof5 dataset, allowing it to learn meaningful representations of audio data without relying on labels. Subsequently, end-to-end training is performed using labeled data, fine-tuning both the feature extractor and the classification head for the specific task. This dual-phase training approach achieves a notable reduction in error rates, resulting in a Dev EER of 20.82\% and an Eval EER of 26.36\%, showcasing the combined benefits of unsupervised pretraining and supervised fine-tuning.

\subsection{MaskedSpec and MaskedFeature Experiments}

\begin{table}[h!]
\centering
\begin{tabular}{@{}lll@{}}
\toprule
Mask Type & Dev EER (\%) & Eval EER (\%) \\
\midrule
None & 2.86 & 6.15 \\
Squares & 1.73 & 6.80 \\
Bands & \textbf{1.36} & \textbf{4.53} \\
Singles & 1.56 & 5.58 \\
Gauss & 3.89 & 6.06 \\
\bottomrule
\end{tabular}

\caption{Model performance comparison of different MaskedSpec masking strategies.}

\label{MaskedSpec_results}
\end{table}

In the following experiment, we evaluate the model's performance when trained with either \textit{MaskedSpec} or \textit{MaskedFeature} augmentation techniques. These experiments are conducted using the XLS-R 53 pertaining for the feature extractor \cite{conneau2020unsupervised}, in combination with compression augmentations. The aim is to assess how the different masking strategies affect the model's ability to generalize and handle artifacts introduced by compression, ultimately improving its performance in detecting deepfake speech.

The results summarized in Table \ref{MaskedSpec_results} highlight the effectiveness of MaskedSpec augmentation with various masking strategies on the model's performance, compared to a baseline model trained without any MaskedSpec augmentation. The \textit{Bands} masking strategy showed best performance, achieving the lowest EERs on both the development set (1.36\%) and the evaluation set (4.53\%), outperforming the baseline in both metrics.

While the \textit{Squares} mask demonstrated a competitive development EER of 1.73\%, its evaluation EER of 6.80\% indicates less consistency when tested on unseen data, suggesting that its improvements are more confined to the training set conditions. The \textit{Singles} mask presented a balanced performance, with moderately low EERs on both the development (1.56\%) and evaluation (5.58\%) sets.
In contrast, the \textit{Gauss} mask exhibited a higher development EER of 3.89\% but still slightly improved the evaluation EER to 6.06\% compared to the baseline. The improvement in evaluation EER over the baseline indicates that the \textit{Gauss} mask still contributes positively to the model's ability to handle real-world distortions, albeit less consistently than strategies such as \textit{Bands} or \textit{Singles}. This highlights the trade-off between overfitting to development conditions and generalizing to unseen data, emphasizing the need to balance augmentation techniques for optimal performance across datasets.

\begin{table}[h!]
\centering
\begin{tabular}{@{}ccc|cc@{}}
 \toprule
 MaskedSpec  & Feature Norm & LPF & Dev EER (\%) & Eval EER (\%) \\

 \midrule
 \multirow{2}{*}{\centering Squares} & \cmark & \xmark & 2.54 & 7.64 \\
                                     & \xmark & \cmark & 1.58 & 6.05 \\
 \midrule
 \multirow{2}{*}{\centering Bands}    & \cmark & \xmark & 1.48 & 5.51 \\
                                     & \xmark & \cmark & 1.74 & 6.75 \\
\midrule
 \multirow{2}{*}{\centering Singles}  & \cmark & \xmark & 2.48 & \textbf{5.25} \\
                                     & \xmark & \cmark & 1.92 & 7.20 \\
 \midrule
 \multirow{2}{*}{\centering Gauss}   & \cmark & \xmark & \textbf{0.94}  & 5.82 \\
& \xmark & \cmark & 2.07 & 5.94 \\
 \bottomrule
\end{tabular}
\caption{Model performance with mixed augmentations, showcasing the effects of various MaskedSpec masking strategies combined with feature normalization and low-pass filtering (LPF).}

\label{augment_experiments}
\end{table}

We further examined the combined effects of MaskedSpec masking strategies with low-pass filtering (LPF) and feature normalization, as shown in Table \ref{augment_experiments}. The \textit{Gauss} mask combined with feature normalization achieved the lowest development EER (0.94\%), demonstrating improved training performance over MaskedSpec alone. This configuration also showed an improvement in evaluation EER, reducing it from 6.06\% without feature normalization to 5.82\%. Conversely, while feature normalization enhanced the evaluation EER for the \textit{Singles} mask, it led to a slight increase in development EER, indicating a trade-off between training and evaluation performance.
Both the \textit{Squares} and \textit{Gauss} masks paired with LPF achieved reduced development and evaluation EERs, highlighting their adaptability to unseen data. However, not all mask and augmentation combinations were equally effective; for example, the \textit{Bands} mask did not yield as robust results, underscoring the need for carefully tailored augmentations to optimize performance across both training and evaluation scenarios.

\begin{table}[h!]
\centering
\begin{tabular}{@{}lll@{}}
\toprule
Mask Type & Dev EER (\%) & Eval EER (\%) \\
\midrule
None & 2.86 & 6.15 \\
 Squares & 2.49 &  6.73 \\
 Bands & 1.44 & 8.78  \\
 Singles & 2.07 & 13.13 \\
 Gauss & \textbf{1.38} & \textbf{5.22} \\
 \bottomrule
\end{tabular}
\caption{Model performance comparison of different MaskedFeature masking strategies.}
\label{feature_masks_results}
\end{table}

In Table \ref{feature_masks_results}, we present the performance of different masking strategies within the \textit{MaskedFeature} augmentation framework compared to the baseline model without masking. All masking strategies outperform the baseline on the development set, demonstrating the effectiveness of \textit{MaskedFeature} in reducing the Dev EER. For instance, the \textit{Gauss} mask achieves the lowest Dev EER of 1.38\%, followed closely by the \textit{Bands} mask at 1.44\%. Similarly, the \textit{Singles} mask and \textit{Squares} mask also show improvements, reducing the Dev EER to 2.07\% and 2.50\%, respectively.

However, the performance on the evaluation set tells a different story. While the \textit{Gauss} mask maintains competitive performance with the lowest Eval EER of 5.22\%. The \textit{Bands} mask, despite its strong Dev EER, sees a significant drop in generalization ability with a high Eval EER of 8.78\%. Similarly, the \textit{Singles} mask, which performed moderately well during development, exhibits poor generalization with an Eval EER of 13.13\%. The \textit{Squares} mask strikes a balance with an Eval EER of 6.73\%, improving over some masks but still falling short in terms of robust generalization. These results highlight that while \textit{MaskedFeature} augmentation effectively reduces the Dev EER, its ability to generalize to unseen data varies significantly across masking types. The \textit{Gauss} mask emerges as the most promising option for balancing development and evaluation performance, indicating a potential for enhancing robustness in practical scenarios.

\subsection{Comparison with Standard Augmentations}

To further assess the effectiveness of our proposed augmentation strategies, we conducted an comprehensive series of experiments employing the most effective configurations of MaskedSpec and MaskedFeature, referred to as $MS_B$ (Bands) and $MF_G$ (Gaussian), respectively.
In these experiments, we examined the performance of our augmentations alongside widely adopted techniques: RawBoost ($R$) and compression ($C$), the latter previously utilized in earlier subsections. These experiments were designed to assess not only the standalone benefit of our augmentations but also their capacity to enhance and complement existing methods.

\begin{table}[h!]
\centering
\footnotesize
\rowcolors{2}{gray!10}{white}
\begin{tabular}{lcc}
\toprule
Augmentation Configuration & Dev EER (\%) & Eval EER (\%) \\
\midrule
None                          & 3.20  & 8.34 \\
\midrule
$C$             & 2.86  & 6.15 \\
$R$                & 5.56  & 6.11 \\
$C$ + $R$                     & 3.51  & 6.14 \\  
\midrule
$MS_B$ + $C$                  & \textbf{1.36}  & \textbf{4.53} \\
$MS_B$ + $R$                  & 2.29  & 4.82 \\
$MS_B$ + $C$ + $R$            & 2.86  & 4.99 \\
\midrule
$MF_G$ + $C$                  & 1.38  & 5.22 \\
$MF_G$ + $R$                  & 3.26  & 5.21 \\
$MF_G$ + $C$ + $R$            & 1.88  & 4.94 \\
\bottomrule
\end{tabular}
\caption{
Equal Error Rates (EER) on the development and evaluation sets under various augmentation configurations. $MS_B$ = MaskedSpec (Bands), $MF_G$ = MaskedFeature (Gaussian)
$C$ = Compression augmentation; $R$ = RawBoost. Best performance was achieved using $MS_B$ + $C$.
}
\label{tab:rawboost_experiments}
\end{table}

Table~\ref{tab:rawboost_experiments} presents model performance across various augmentation configurations. As a baseline, the model trained without any augmentation achieves a relatively high EER, particularly on the evaluation set (8.34\%). Applying compression alone reduces the EER to 6.15\%, while RawBoost achieves a comparable result (6.11\%) but with a weaker Dev EER (5.56\%). Interestingly, combining both compression and RawBoost does not yield further improvement, indicating potential redundancy or saturation effects between these augmentations.

In contrast, when our MaskedSpec ($MS_B$) or MaskedFeature ($MF_G$) augmentations are introduced, consistent gains are observed. Specifically, the combination of compression with $MS_B$ reduces Eval EER to 4.53\%, while adding $MS_B$ to RawBoost results in a strong 4.82\%. The same trend holds for $MF_G$, which lowers Eval EER to 5.22\% and 5.21\% when combined with compression and RawBoost, respectively. Notably, the combination $MF_G$ + $C$ + $R$ achieves better result at 4.94\%, further highlighting the compatibility of our augmentation with existing techniques.

From a theoretical perspective, these improvements can be attributed to the nature of the distortions introduced. While RawBoost targets temporal variability by modifying amplitude and phase characteristics, and compression introduces global artifacts due to codec degradation, our augmentations inject structured masking along the frequency or feature axis. This masking enforces robustness to localized distortions and promotes better generalization across unseen manipulations. Thus, our methods operate on orthogonal principles relative to $C$ and $R$, providing complementary benefits when combined.

\begin{table}[h!]
\centering
\footnotesize
\begin{tabular}{@{}cccccc|c@{}}
\toprule
$MS_B$ + $C$ & $MS_B$ + $R$ & $MS_B$ + $C$ + $R$ & $MF_G$ + $C$ & $MF_G$ + $R$ & $MF_G$ + $C$ + $R$ & Eval EER (\%) \\
\midrule
\cmark & \cmark & \cmark & \cmark & \cmark & \cmark & \textbf{3.74} \\
\midrule
\cmark & \cmark & \xmark & \cmark & \cmark & \xmark & 4.02 \\
\cmark & \xmark & \cmark & \cmark & \xmark & \cmark & 3.82 \\
\xmark & \cmark & \cmark & \xmark & \cmark & \cmark & 3.98 \\
\midrule
\cmark & \cmark & \cmark & \xmark & \xmark & \xmark & 3.92 \\
\xmark & \xmark & \xmark & \cmark & \cmark & \cmark & 4.01 \\
\bottomrule
\end{tabular}
\caption{
Fusion performance (Eval EER \%) for models trained with different augmentation configurations. $MS_B$ and $MF_G$ denote MaskedSpec and MaskedFeature, respectively. $C$ = Compression augmentation, $R$ = Rawboost. Each row represents a distinct combination of fusion sources. The best performance achieved when fusing all six augmentation configurations.
}
\label{tab:fusion-augments}
\end{table}

In addition to evaluating individual augmentation effects, we explore the impact of combining models trained under different augmentation settings via late score-level fusion. Each fusion configuration in Table~\ref{tab:fusion-augments} is composed of the same models and training process, trained with the respective augmentations indicated in the columns. The goal is to assess whether diverse augmentation conditions lead to complementary decision patterns that enhance overall detection performance when combined.

As shown, the full fusion of all augmentations achieves the lowest EER (3.74\%), indicating strong complementarity among the models. Removing any individual augmentation consistently degrades performance. This suggests that each component introduces unique perturbations that contribute valuable diversity in the learned feature space. The results highlight the benefit of leveraging augmentation diversity not only during training, but also in inference via ensemble-based strategies.

\subsection{Different Feature Extractor Pretraining}

Pretrained initialization of the feature extractor plays a significant role in shaping how speech signals are represented during downstream training, particularly in end-to-end architectures. Differences in dataset scale, language diversity, and domain alignment of the pretraining corpus can lead to varied representational biases, ultimately influencing performance. To examine these effects, we evaluate several feature extractor initializations using a unified training pipeline that incorporates our proposed augmentation strategies: MaskedSpec with Band masking ($MS_B$), MaskedFeature with Gaussian masking ($MF_G$), and compression augmentations.

\begin{table}[h!]
\centering
\footnotesize
\begin{tabular}{@{}cccccc@{}}
 \toprule
 Feature Extractor Pretraining & Compressions & $MS_B$ & $MF_G$ & Dev EER (\%) & Eval EER (\%) \\
 \midrule

 \multirow{2}{*}{XLS-R 53} & \xmark & \xmark & \xmark & 3.20 & 8.34 \\
  & \cmark & \xmark & \xmark & 2.86 & 6.15  \\

 \midrule

 \multirow{2}{*}{ASVSpoof5} & \xmark & \xmark & \xmark & 20.82 & 26.36 \\
   & \cmark & \xmark & \xmark & 16.03 & 20.51 \\
 
\midrule
\midrule

 \multirow{2}{*}{XLS-R 53} 
& \cmark & \cmark & \xmark & 1.36 &  4.53 \\
& \cmark & \xmark & \cmark & 1.38 & 5.22 \\

 \midrule
 
 \multirow{2}{*}{XLS-R 128}  & \cmark & \cmark & \xmark & 0.34 & 4.74 \\
  & \cmark & \xmark & \cmark & \textbf{0.29} & 5.01 \\

 \midrule
 
 \multirow{2}{*}{Large-960h} & \cmark & \cmark & \xmark & 6.03 & 6.05  \\
  & \cmark & \xmark & \cmark & 9.75 & 8.25 \\
 
\midrule

 \multirow{2}{*}{WAVLM} & \cmark & \cmark & \xmark & 0.86 & 4.77 \\
  & \cmark & \xmark & \cmark & 2.57 & 6.96 \\
 
\midrule

 \multirow{2}{*}{ASVSpoof5} & \cmark & \cmark & \xmark & 12.49 & 12.66 \\
  & \cmark & \xmark & \cmark & 17.03 & 14.17 \\

\midrule

 \multirow{2}{*}{ASVSpoof5+}  & \cmark & \cmark & \xmark & 5.12 & \textbf{4.44} \\
 & \cmark & \xmark & \cmark & 5.62 & 5.27 \\

\bottomrule
\end{tabular}
\caption{Performance comparison across pretrained feature extractors and augmentation strategies. $MS_B$ = MaskedSpec (Bands), $MF_G$ = MaskedFeature (Gaussian), and $C$ = Compression augmentation. Best evaluation EER scores are highlighted in bold.}
\label{wav2vec_pretrain_results}
\end{table}

Table~\ref{wav2vec_pretrain_results} summarizes the performance of each model configuration. As a baseline, we assess models initialized with \textit{XLS-R 53} and \textit{ASVSpoof5}, both trained without any augmentation. \textit{XLS-R 53}, pretrained on over 50k hours of multilingual speech, offers a broad and diverse representation space. In contrast, the \textit{ASVSpoof5} pretrained weights, although task-specific and focused on spoofed and bona fide speech, is limited in both scale and language diversity. This constraint limits its capacity to produce generalizable embeddings, resulting in significantly higher error rates.
We then apply our augmentation strategy across all initialization settings. For \textit{XLS-R 53}, incorporating $MS_B$ leads to a 26.4\% reduction in Eval EER, while $MF_G$ results in a 15.1\% decrease. The improvements are even more substantial for \textit{ASVSpoof5}, with $MS_B$ reducing the Eval EER from 20.51\% to 12.66\% (38.3\% improvement), and $MF_G$ from 20.51\% to 14.17\% (30.9\% improvement). 

The \textit{ASVSpoof5+} configuration introduces an additional insight, where compression augmentation applied during pretraining. This technique simulates real-world distortions and helps mitigate the limitations of low-resource datasets. With \textit{ASVSpoof5+}, Eval EER is further reduced to 4.44\% with $MS_B$ and 5.27\% with $MF_G$, corresponding to relative improvements of 64.9\% and 62.8\% compared to the standard \textit{ASVSpoof5} configuration. This highlighting the impact of early exposure to distortion in enhancing feature learning.

To broaden our analysis, we also examine several publicly available large-scale models. \textit{XLS-R 128}, trained on 436k hours across 128 languages, achieves the best overall results for Dev EER with 0.29\% with $MF_G$, and one of the best Eval EERs (4.74\%) with $MS_B$. These outcomes underscore the benefit of wide linguistic and acoustic coverage in self-supervised pretraining. \textit{WavLM}, although monolingual, is trained on 94k hours of English speech and delivers competitive performance, suggesting that scale alone can provide strong generalization capabilities when paired with effective augmentations. In contrast, the Large-960h model, trained on 960 hours of LibriSpeech, underperforms relative to other models, illustrating the limitations of narrow, domain-specific pretraining.

The full set of results, including experiments with RawBoost and evaluation with minDCF, are presented in the Appendix.\footnote{Please refer to Appendix~\ref{appendix:rawboost-minDCF} for detailed results.}

\subsection{Fusion of Models}

\begin{table}[h!]
\centering
\footnotesize
\rowcolors{2}{gray!10}{white}

\resizebox{\textwidth}{!}{%
\begin{tabular}{l|cc|cc|cc|cc|c}
\toprule
Fusion Config & \multicolumn{2}{c|}{ASVSpoof5+} & \multicolumn{2}{c|}{XLS-R 53} & \multicolumn{2}{c|}{XLS-R 128} & \multicolumn{2}{c|}{WavLM} & Eval EER (\%) \\
\cmidrule{2-9}
 & $MS_B$ & $MF_G$ & $MS_B$ & $MF_G$ & $MS_B$ & $MF_G$ & $MS_B$ & $MF_G$ & \\
\midrule
$C$ only     & \cmark & \cmark &        &        &        &        &        &        & 4.08 \\
\midrule
$C$ only                        & \cmark &        & \cmark &        & \cmark &        & \cmark &        & 3.53 \\
$C$ only                        &        & \cmark &        & \cmark &        & \cmark &        & \cmark & 3.69 \\
$C$ only                        & \cmark & \cmark & \cmark & \cmark & \cmark & \cmark & \cmark & \cmark & 3.19 \\
\midrule
$C$, $R$, $C{+}R$ & \cmark* &        & \cmark &        & \cmark &        & \cmark &        & 2.92 \\
$C$, $R$, $C{+}R$ &        & \cmark* &        & \cmark &        & \cmark &        & \cmark & 3.14 \\
$C$, $R$, $C{+}R$ (full fusion)    & \cmark* & \cmark* & \cmark & \cmark & \cmark & \cmark & \cmark & \cmark & \textbf{2.71} \\
\bottomrule
\end{tabular}
}
\caption{
\textbf{Fusion performance} across models trained with different augmentation combinations and pretrained feature extractors.  
Augmentations used include compression ($C$), RawBoost ($R$), and their combination ($C+R$).  
\textbf{*} Models pretrained with RawBoost on ASVSpoof5+ were excluded due to degraded performance. The best fusion result is achieved when combining all models.
}
\label{tab:fusion_results}
\end{table}

Table~\ref{tab:fusion_results} presents the evaluation results of fusion across models trained with different augmentation configurations and feature extractor initializations. The fusion is performed by averaging output scores from independently trained models. This analysis examines whether combining models trained under complementary augmentation strategies leads to improved detection performance.

We first explore fusion within a closed condition setting, using only models trained with ASVSpoof5 dataset. When both masking strategies ($MS_B$ and $MF_G$) are included, the fused system achieves an Eval EER of 4.08\%, surpassing the performance of individual models in isolation.

We then expand the fusion scope to include models trained with different feature extractor initializations (\textit{XLS-R 53}, \textit{XLS-R 128},\textit{ WavLM}), trained with compression. Fusing models trained individually with either $MS_B$ or $MF_G$ produces notable improvements (Eval EER of 3.53\% and 3.69\%, respectively). However, combining both masking strategies across all extractors further improves performance, yielding an Eval EER of 3.19\%. 

Finally, we extend the fusion to encompass models trained with compression, RawBoost, and their combination. While RawBoost had degraded results under \textit{ASVSpoof5+} and is excluded in those cases (marked with *), its inclusion from other pretraining sources contributes valuable temporal perturbation patterns. Fusion of models trained only with $MS_B$ or only with $MF_G$ gives Eval EERs of 2.92\% and 3.14\%, respectively. The full combination of both masking types across all augmentation regimes achieves the lowest error rate of 2.81\%.

\subsection{Cross-Dataset Evaluation}
To rigorously assess the generalization of our framework under domain shift, we conduct cross-dataset evaluations and benchmark against leading systems reported in \cite{li2025survey}. While most prior systems were trained on ASVSpoof2019, our primary configuration was trained on ASVSspoof5 (2024) (Track~1). This comparison highlights the effect of different training sets when models are evaluated beyond their source domain.

\begin{table}[h!]
\footnotesize
\centering
\begin{tabular}{@{}lccccc@{}}
\toprule
System & Training Data & ASVSpoof19 Eval & ASVSpoof21 LA & ASVSpoof21 DF \\
\midrule
AASIST + RawBoost  & ASVSpoof2019  & - & \cellcolor{gray!60}0.82 & \cellcolor{gray!60}2.85 \\
SDC + BiLSTM       & ASVSpoof2019  & \cellcolor{gray!30}0.22 & \cellcolor{gray!30}3.50 & 3.41 \\
Rawformer          & ASVSpoof2019  & 0.59 & 4.98 & 4.53 \\
Bi-LSTM + MLP      & ASVSpoof2019  & 1.28 & 6.53 & 4.75 \\
DARTS              & ASVSpoof2019  & 1.08 & --   & 7.89 \\
\midrule
\textbf{Ours}      & ASVSpoof5 (2024)  & 4.74 & 12.58 & 7.94 \\
\midrule
\textbf{Ours}      & ASVSpoof2019  &\cellcolor{gray!60}0.18 & 5.25 & \cellcolor{gray!30}2.92 \\
\bottomrule
\end{tabular}
\caption{EER (\%) comparison on ASVSpoof2019 evaluation and ASVSpoof2021 LA and DF evaluation sets. The training dataset used for each system is explicitly indicated. Dark gray indicates the best-performing system, and light gray denotes the second-best.}
\label{tab:eer_with_training}
\end{table}

Table~\ref{tab:eer_with_training} shows that when trained on ASVSpoof5 (2024), our system demonstrates reasonable resilience but does not reach state-of-the-art performance under cross-dataset evaluation. While these results are acceptable, we sought to determine whether the limitation stems from the model itself or from dataset mismatch. To this end, we trained the system on the ASVSpoof2019 training set, which is considerably smaller than ASVSpoof5, and compared it against current state-of-the-art systems.
In this setting, our model achieved second-best performance on both the ASVSpoof2019 evaluation and the ASVSpoof2021 DF tasks, with EERs of 0.18\% and 2.92\%, respectively. These results underscore the effectiveness of our augmentation framework even when trained on a substantially smaller dataset.

From these results, we conclude that:  
1. The relatively weaker performance of ASVSpoof5-trained models is primarily due to dataset mismatch rather than inherent shortcomings of our framework.  
2. Resilience to domain shift remains a central challenge for practical deployment. We believe that designing feature extractors capable of learning general and invariant representations is key to improving robustness and ensuring reliability in industrial systems.

\section{Analysis of Model Performance}
In this section, we present a detailed analysis of our best preformed systems using the evaluation methods provided by the ASVspoof5 challenge organizers. We examine aggregated trends in EERs across a range of spoofing attacks  types and codecs conditions, allowing us to infer model behavior. By analyzing pooled EERs per condition, we could identify patterns of vulnerability and resilience, assess the effectiveness of our augmentation strategies, and highlight which model configurations offer the most consistent performance across diverse real-world scenarios.
\subsection{Spoofing Attacks}

\begin{table}[h!]
\centering
\scriptsize
\begin{tabular}{cp{0.3cm}p{0.3cm}p{0.3cm}p{0.3cm}p{0.3cm}p{0.3cm}p{0.3cm}p{0.3cm}p{0.3cm}p{0.3cm}p{0.3cm}p{0.4cm}p{0.3cm}p{0.3cm}p{0.3cm}p{0.3cm}}
 \hline
 Model & A17 & A18 & A19 & A20 & A21 & A22 & A23 & A24 & A25 & A26 & A27 & A28 & A29 & A30 & A31 & A32 \\
 \toprule

 XLSR53, $MS_B$ + C &1.62& 2.68& \textbf{1.32}& \textbf{1.33}& 1.56& 2.34& 3.53& 6.03& \textbf{1.42}& \textcolor{lightgray}{1.41}& 1.76& 17.32& 1.42& 3.76& 5.10& \textbf{1.47}\\
XLSR53, $MF_G$ + C&2.00& 4.53& 2.01& 1.75& 2.22& 2.58& 5.05& 8.71& 1.63& 1.94& 2.09& 16.30& 1.74& 5.52& 6.60& \textcolor{lightgray}{1.55}\\

XLSR128, $MS_B$ + R& \textcolor{lightgray}{0.96}& 2.77& 2.33& 2.36& \textcolor{lightgray}{0.92}& \textcolor{lightgray}{1.46}& 2.99& \textcolor{lightgray}{3.06}& 1.55& 1.50& 5.12& \textcolor{lightgray}{8.99}& \textcolor{lightgray}{0.75}& 7.01& 6.34& 3.57\\
XLSR128, $MS_B$ + R + C  &\textbf{0.71}& 2.15& 2.43& 3.70& \textbf{0.66}& \textbf{1.35}& 2.91& \textbf{3.04}& \textcolor{lightgray}{1.54}& \textcolor{lightgray}{1.41}& 6.34& \textbf{6.05}& \textbf{0.58}& 6.63& 6.50& 6.12\\
XLSR128, $MF_G$ +  C  &1.86& 4.18& 1.79& \textcolor{lightgray}{1.48}& 1.14& 1.76& 2.61& 5.81& 1.73& 1.81& 3.60& 12.74& 1.42& 6.96& 6.46& 2.47\\
XLSR128, $MF_G$ + R +   C  &1.05& 2.77& \textcolor{lightgray}{1.76}& 2.92& 1.18& 1.50& 4.67& 3.58& 1.61& 1.63& 5.50& 9.18& 0.77& 6.41& 7.53& 3.96\\

ASVSpoof5, $MS_B$ + C &1.69& \textcolor{lightgray}{0.84}& 6.20& 3.42& 2.29& 3.10& \textcolor{lightgray}{1.63}& 4.55& 3.83& 1.45& \textbf{1.25}& 16.14& 1.08& \textbf{1.19}& \textbf{3.10}& 1.62\\
ASVSpoof5, $MF_G$ +  C  &2.02& \textbf{0.82}& 6.49& 5.34& 1.85& 4.30& \textbf{1.47}& 5.57& 5.20& \textbf{1.37}&  \textcolor{lightgray}{1.34}& 18.22& 1.14& \textcolor{lightgray}{1.49}& \textcolor{lightgray}{4.60}& 2.13\\

\midrule
\midrule

ASVSpoof5+ fusion &1.16& 0.58& 5.23& 3.51& 1.53& 2.66& 1.19& 4.24& 3.34& 1.22& 1.13& 15.01& 0.83& 0.94& 2.94& 1.56\\
Full fusion &0.55& 0.94& 1.58& 1.58& 0.89& 1.18& 1.68& 3.53& 0.91& 1.06& 1.82& 7.86& 0.54& 1.88& 3.33& 1.49\\

 \bottomrule
\end{tabular}
\caption{Pooled EER performance analysis results for a selected subset of top-performing models across 16 spoofing attack types (A17–A32). \textbf{Bold} values indicate the lowest EER for each codec among the single models, while \textcolor{lightgray}{light gray} denotes the second-best. }

\label{analysis_results}
\end{table}

The ASVSpoof5 evaluation set introduces a diverse range of unseen spoofing attacks, labeled A17–A32. Table~\ref{analysis_results} presents the pooled EERs for each attack across various model configurations. As observed, attacks such as A17, A21, A22, A25, A26, and A29 consistently result in low EERs across most models, suggesting that these spoofing types contain identifiable artifacts or exhibit acoustic patterns that are effectively addressed by the applied augmentations. Attacks A24, A28, A30, and A31 present more substantial challenges, as evidenced by elevated error rates in nearly all configurations. Among them, A28 stands out as the most difficult to detect, with numerous models exceeding 10\% EER, indicating its high perceptual similarity to bona fide speech.

Model initialized with \textit{XLS-R 53} and trained using $MS_B$ + $C$ exhibit strong generalization, achieving low EERs on a diverse set of attacks, including A19, A20, A25, and A32. In particular, \textit{XLS-R 128} combined with $MS_B$ + $R$ + $C$ delivers leading performance across both moderate and high-difficulty spoofing attacks scenarios, including A24 and A28. This supports the value of leveraging extensive multilingual pretraining along with complementary augmentations for enhanced robustness.

Interestingly, models pretrained on the more constrained \textit{ASVSpoof5+} dataset outperform larger models on several spoofing types where general-purpose pretraining appears less effective. Specifically, the $MS_B$ + $C$ variant achieves the lowest EER on A27, A30, and A31, while $MF_G$ + $C$ outperforms others on A18, A23, and A26. These outcomes indicate that domain-specific pretraining, even at smaller scale, can offer competitive advantages when paired with task-relevant augmentations, likely due to improved alignment with the spectral and temporal characteristics of fake audio.

\subsection{Codecs}

\begin{table}[h!]
\centering
\scriptsize
\begin{tabular}{cp{0.8cm}p{0.6cm}p{0.6cm}p{0.6cm}p{0.6cm}p{0.6cm}p{0.6cm}p{0.6cm}p{0.6cm}p{0.6cm}p{0.6cm}p{0.6cm}}
 \hline
 Model & None & C01& C02& C03& C04& C05& C06& C07& C08& C09& C10& C11 \\
 \toprule
XLSR128, $MS_B$ + C & \textbf{0.59}& \textbf{2.17}& \textcolor{lightgray}{2.14}& 3.62& 12.13& \textbf{0.72}& \textbf{1.18}& 16.29& 5.43& \textbf{3.61}& \textcolor{lightgray}{5.04}& \textbf{1.06}\\
XLSR128, $MS_B$ + R &1.44& 2.62& 2.43& \textcolor{lightgray}{3.60}& 8.59& 1.67& 2.12& 11.03& \textcolor{lightgray}{4.37}& \textcolor{lightgray}{3.80}& 5.20& 1.70\\
XLSR128, $MS_B$ + R + C&1.10& \textcolor{lightgray}{2.54}& 3.08& \textbf{3.35}& 9.17& 1.52& \textcolor{lightgray}{1.77}& 12.51& \textbf{4.13}& 4.02& 5.60& \textcolor{lightgray}{1.21}\\
XLSR128, $MF_G$ + R+ C &1.84& 2.89& 3.21& 4.65& 8.31& 2.59& 2.56& 11.07& 4.58& 4.05& \textbf{4.60}& 1.83\\

WAVLM, $MS_B$+R+C & \textcolor{lightgray}{1.00}& 3.66& \textbf{2.11}& 4.12& 9.27& \textcolor{lightgray}{1.25}& 2.48& 12.77& 4.55& 5.53& 6.17& 1.63\\

ASVSpoof5, $MS_B$ + C &3.35& 4.09& 3.62& 5.07& \textbf{4.97}& 4.13& 4.29& \textbf{5.85}& 5.87& 4.76& 5.55& 3.42\\
ASVSpoof5, $MF_G$ + C &3.83& 4.83& 4.44& 5.98& \textcolor{lightgray}{6.05}& 5.19& 4.92& \textcolor{lightgray}{7.09}& 7.24& 6.33& 6.65& 3.82\\

 




\midrule
\midrule
AVSpoof5+ fusion &2.98& 3.61& 3.28& 4.64& 4.68& 4.09& 4.04& 5.64& 5.34& 4.57& 5.15& 2.98\\
Best fusion &0.44& 1.39& 0.95& 2.37& 6.04& 0.62& 0.97& 8.21& 2.87& 2.19& 3.30& 0.90\\

 \bottomrule
\end{tabular}
\caption{Pooled EER performance analysis for a selected subset of top-performing models across 12 codec conditions (C01–C11) and the uncompressed baseline (“None”). \textbf{Bold} values indicate the lowest EER for each codec among the single models, while \textcolor{lightgray}{light gray} denotes the second-best. }
\label{codec_analysis_results}
\end{table}

Examining detection difficulty across codecs revealed that C04 and C07 presented particular challenges, yielding consistently high EERs across models. This suggests that these codecs may mask or distort essential audio features that models rely on. Conversely, C01, C02, C05, C06, C11, and None (uncompressed) audio exhibited lower EERs, indicating that models were generally able to process and identify fake audio more effectively within these formats. 

The results presented in Table \ref{codec_analysis_results} reveal that models pretrained on large and diverse corpora (\textit{XLS-R 128} and \textit{WavLM}) consistently achieve lower EERs. Augmentation with MaskedSpec ($MS_B$) and compression ($C$) proves particularly effective across a wide range of codecs. For example, the \textit{XLS-R 128} model with $MS_B$ + C achieves the best performance on codecs such as None, C01, C06, C09, and C11, and maintains competitive results elsewhere.

Despite its limited data scale, the \textit{ASVSpoof5+} demonstrates notable improvements. While it underperforms compared to large-scale pretrained models on average, it achieves the lowest EERs in certain challenging codecs where XLS-R 128 struggles, such as C04 and C07. This indicates that task-specific distortions seen during \textit{ASVSpoof5+} pretraining may offer benefits for generalization under similar conditions.

\subsection{Analysis-Driven Fusion Strategy}

The analysis reveals that several models exhibit complementary strengths across different spoofing attacks and codec conditions. Notably, the models \textit{XLS-R 128} with $MS_B$ + $R$ + $C$, \textit{XLS-R 53} with $MS_B$ + $C$, and \textit{ASVSpoof5+} with both $MS_B$ + $C$ and $MF_G$ + $C$ demonstrate strong yet distinct performance profiles. 

Although none of these models individually achieved the best performance on either the development or evaluation sets, their fusion leads to the lowest observed EER of 2.58\% (Table~\ref{tab:fusion_analysis_results}). This highlights a critical insight: the effectiveness of model fusion lies not in combining top-performing models alone, but in leveraging the complementary error patterns and inductive biases across architectures and augmentations. Fusion strategies motivated by analytical diversity rather than raw accuracy can yield more robust and generalizable systems.

\begin{table}[h!]
\centering
\footnotesize
\begin{tabular}{l|c}
\toprule
Fused Models & ASVSpoof5 Eval EER (\%) \\
\midrule
XLS-R 128, $MS_B$ + $R$ + $C$ \\
XLS-R 53, $MS_B$ + $C$ \\
ASVSpoof5+, $MS_B$ + $C$ \\
ASVSpoof5+, $MF_G$ + $C$ & \multirow{-4}{*}{\textbf{2.58}} \\
\bottomrule
\end{tabular}
\caption{
Fusion of four strategically selected models on the ASVspoof5 evaluation set, chosen for their complementary performance under codec and spoofing attack conditions. Although none are the top individual performers, their diverse strengths yield the lowest overall evaluation EER, demonstrating the benefit of cross-extractor, cross-augmentation fusion.
}
\label{tab:fusion_analysis_results}
\end{table}

\section{Discussion}

\subsection{Spectral and Feature-Level Masking}
We evaluate four masking strategies at both the spectral and feature levels. The comparison of mask shapes within MaskedSpec and MaskedFeature (Tables~\ref{MaskedSpec_results} and~\ref{feature_masks_results}) identifies the most effective configurations: \textit{Bands} for MaskedSpec and \textit{Gauss} for MaskedFeature. 

From a theoretical perspective, these masking strategies can be viewed as structured forms of dropout, tailored to the audio domain. Spectral \textit{Bands} masking acts as dropout in the frequency domain, occluding contiguous frequency regions and compelling the model to infer missing spectral content from global context rather than relying on narrowband artifacts. Feature-level \textit{Gauss} masking operates within the latent representation space, where attenuating localized activations while preserving global structure promotes the development of distributed, noise tolerant embeddings. This effect mirrors observations in masked image modeling, where structured masking has been shown to encourage context-aware and generalizable representations.

These strategies provide complementary benefits to RawBoost and compression augmentations (Table~\ref{tab:rawboost_experiments}), and their joint use yields consistent improvements, as reflected in the ASVSpoof5+ fusion results (Table~\ref{tab:fusion_results}). Importantly, the effectiveness of these masking methods remains stable across different pretraining conditions, highlighting their robustness and general applicability.

\subsection{Augmentation in Self-Supervised Pretraining}

To enhance representation learning under limited-domain conditions, we applied compression augmentation during the self-supervised pretraining phase of ASVSpoof5+. Our goal was to encourage the development of invariant features capable of generalizing across degraded acoustic conditions. The effect of this augmentation was substantial and led to over 60\% relative improvment.

Interestingly, models trained with RawBoost during the end-to-end phase, but without exposure to it in pretraining, consistently underperformed. The RawBoost-augmented ASVSpoof5+ configurations yielded poor results and were ultimately excluded from our fusion ensemble. This suggests that certain augmentations may need to be introduced earlier in the training pipeline to fully integrate their distortive priors into the learned representation. 

Notably, despite its smaller scale and limited linguistic diversity, the ASVSpoof5+ initialization was able to match or exceed the performance of much larger, general-purpose initializations such as XLS-R 128 and WavLM. This highlights the effectiveness of domain-specific augmentation strategies in improving self-supervised learning under constrained data regimes.

\subsection{Fusion of Diverse Models and Augmentation Strategies}

Another key outcome of our study is the benefit of combining models trained with different pretraining initializations and augmentation strategies. Pretrained feature extractors such as XLS-R 128 and ASVSpoof5+ encode speech representations shaped by the scale, language diversity, and domain specificity of their respective training corpora. Moreover, our augmentation techniques, includes augmentation applied at both the raw audio and feature levels, introduce varied perturbations that promote resilience to distributional shifts and unseen spoofing attacks.

Fusion proved to be an effective strategy for enhancing system robustness, resulting in improved performance under previously unseen distortions and manipulations. As demonstrated in our fusion experiments, combining models trained with both raw audio and feature-level augmentations, as well as using diverse pretraining configurations, reduced the evaluation EER to 2.81\%, significantly outperforming any individual model. These results highlight the importance of leveraging complementary model characteristics, indicating that diversity in training settings contributes meaningfully to the development of resilient and generalizable systems.

\section{Conclusion and Future Work}

This study presents a comprehensive framework for deepfake speech detection that integrates a multi-stage augmentation strategy with hybrid training. We propose novel masking augmentations applied at both the raw audio spectrogram and  learned representations, and conduct an empirical analysis of their effects across different masks shapes. Our findings indicate that \textit{Bands} masking on spectrograms and \textit{Gauss} masking on feature level are particularly effective, leading to consistent reductions in error rates across multiple model backbones.

Additionally, we introduce a targeted compression augmentation strategy applied during the self-supervised pretraining phase, addressing the vulnerability of self-supervision in low-resource scenarios. Introducing this augmentation resulted in a reduction of over 60\% in EER and contributed to improved convergence during end-to-end training.

Our hybrid training framework combines self-supervised pretraining of the feature extractor with a ResNet-34 classification head, which are subsequently trained jointly end-to-end. On the ASVspoof5 (Track 1) benchmark, our system achieved state-of-the-art performance under the closed condition. Furthermore, when evaluated on ASVspoof2019 and ASVspoof2021 DF, the framework reached 0.18\% and 2.92\% EER respectively. 

While our approach achieves leading results on multiple benchmarks, we believe future work should focus on creating cross-domain unified feature representations to further strengthen performance in real-world scenarios. Improving generalization across diverse audio sources remains a key challenge, and advancing feature extractors that learn consistent, domain-invariant representations will be critical for deploying reliable deepfake detection systems in practice.

\bibliographystyle{elsarticle-num}
\bibliography{references}

\begin{thebibliography}{10}
\expandafter\ifx\csname url\endcsname\relax
  \def\url#1{\texttt{#1}}\fi
\expandafter\ifx\csname urlprefix\endcsname\relax\def\urlprefix{URL }\fi
\expandafter\ifx\csname href\endcsname\relax
  \def\href#1#2{#2} \def\path#1{#1}\fi

\bibitem{mcuba2023effect}
M.~Mcuba, A.~Singh, R.~A. Ikuesan, H.~Venter, The effect of deep learning methods on deepfake audio detection for digital investigation, Procedia Computer Science 219 (2023) 211--219.
\newblock \href {https://doi.org/10.1016/j.procs.2023.01.283} {\path{doi:10.1016/j.procs.2023.01.283}}.

\bibitem{Sahidullah2015}
M.~Sahidullah, T.~Kinnunen, A comparison of features for synthetic speech detection, Interspeech (2015) 2087--2091\href {https://doi.org/10.21437/Interspeech.2015-472} {\path{doi:10.21437/Interspeech.2015-472}}.

\bibitem{yi2023audio}
J.~Yi, C.~Wang, J.~Tao, X.~Zhang, C.~Y. Zhang, Y.~Zhao, Audio deepfake detection: A survey, arXiv preprint arXiv:2308.14970 (2023).
\newblock \href {https://doi.org/10.48550/arXiv.2308.14970} {\path{doi:10.48550/arXiv.2308.14970}}.

\bibitem{alzantot2019deep}
M.~Alzantot, Z.~Wang, M.~B. Srivastava, Deep residual neural networks for audio spoofing detection, arXiv preprint arXiv:1907.00501 (2019).
\newblock \href {https://doi.org/10.21437/Interspeech.2019-3174} {\path{doi:10.21437/Interspeech.2019-3174}}.

\bibitem{Rimon2022}
A.~Cohen, I.~Rimon, E.~Aflalo, H.~Permuter, \href{https://www.sciencedirect.com/science/article/pii/S0167639322000619}{A study on data augmentation in voice anti-spoofing}, in: Proceedings of the Annual Conference of the International Speech Communication Association (INTERSPEECH), 2021, pp. 1234--1238.
\newblock \href {https://doi.org/10.1016/j.specom.2022.04.005} {\path{doi:10.1016/j.specom.2022.04.005}}.
\newline\urlprefix\url{https://www.sciencedirect.com/science/article/pii/S0167639322000619}

\bibitem{tak2022automatic}
H.~Tak, M.~Todisco, X.~Wang, J.-w. Jung, J.~Yamagishi, N.~Evans, Automatic speaker verification spoofing and deepfake detection using wav2vec 2.0 and data augmentation, arXiv preprint arXiv:2202.12233 (2022).
\newblock \href {https://doi.org/10.48550/arXiv.2202.12233} {\path{doi:10.48550/arXiv.2202.12233}}.

\bibitem{ASVSpoof2019}
M.~Todisco, X.~Wang, V.~Vestman, M.~Sahidullah, H.~Delgado, A.~Nautsch, J.~Yamagishi, N.~Evans, T.~Kinnunen, K.~A. Lee, Asvspoof 2019: Future horizons in spoofed and fake audio detection, arXiv preprint arXiv:1904.05441 (2019).
\newblock \href {https://doi.org/10.48550/arXiv.1904.05441} {\path{doi:10.48550/arXiv.1904.05441}}.

\bibitem{ASVSpoof2021}
J.~Yamagishi, X.~Wang, M.~Todisco, M.~Sahidullah, J.~Patino, A.~Nautsch, X.~Liu, K.~A. Lee, T.~Kinnunen, N.~Evans, et~al., Asvspoof 2021: accelerating progress in spoofed and deepfake speech detection (2021).
\newblock \href {https://doi.org/10.48550/arXiv.2109.00537} {\path{doi:10.48550/arXiv.2109.00537}}.

\bibitem{ASVSpoof5}
J.-w. J. T. K. I. K. K. A. L. X. L. H.-j. S. M. S. H. T. M. T. X. W. J.~Y. H´ector~Delgado, Nicholas~Evans, \href{https://www.asvspoof.org/file/ASVspoof5___Evaluation_Plan_Phase2.pdf}{Asvspoof 5 evaluation plan} (2024).
\newline\urlprefix\url{https://www.asvspoof.org/file/ASVspoof5___Evaluation_Plan_Phase2.pdf}

\bibitem{oppenheim1999discrete}
A.~V. Oppenheim, R.~W. Schafer, Discrete-time signal processing, Prentice-Hall, 1999.

\bibitem{mittal2022automatic}
A.~Mittal, M.~Dua, Automatic speaker verification systems and spoof detection techniques: review and analysis, International Journal of Speech Technology 25~(1) (2022) 105--134.
\newblock \href {https://doi.org/10.1007/s10772-021-09876-2} {\path{doi:10.1007/s10772-021-09876-2}}.

\bibitem{li2024audio}
M.~Li, Y.~Ahmadiadli, X.-P. Zhang, Audio anti-spoofing detection: A survey, arXiv preprint arXiv:2404.13914 (2024).
\newblock \href {https://doi.org/10.48550/arXiv.2404.13914} {\path{doi:10.48550/arXiv.2404.13914}}.

\bibitem{baevski2020wav2vec}
A.~Baevski, Y.~Zhou, A.~Mohamed, M.~Auli, wav2vec 2.0: A framework for self-supervised learning of speech representations, Advances in neural information processing systems 33 (2020) 12449--12460.
\newblock \href {https://doi.org/arXiv.2006.11477} {\path{doi:arXiv.2006.11477}}.

\bibitem{chen2022wavlm}
S.~Chen, C.~Wang, Z.~Chen, Y.~Wu, S.~Liu, L.~Xie, M.~Zhou, Wavlm: Large-scale self-supervised pre-training for full stack speech processing, in: Proceedings of the 22nd Annual Conference of the International Speech Communication Association (INTERSPEECH), 2022, pp. 3615--3619.

\bibitem{xie2021siamese}
Y.~Xie, Z.~Zhang, Y.~Yang, Siamese network with wav2vec feature for spoofing speech detection., in: Interspeech, 2021, pp. 4269--4273.
\newblock \href {https://doi.org/10.21437/Interspeech.2021-847} {\path{doi:10.21437/Interspeech.2021-847}}.

\bibitem{zeng2023deepfake}
X.-M. Zeng, J.-T. Zhang, K.~Li, Z.-L. Liu, W.-L. Xie, Y.~Song, \href{https://ceur-ws.org/Vol-3597/paper6.pdf}{Deepfake algorithm recognition system with augmented data for add 2023 challenge.}, in: DADA@ IJCAI, 2023, pp. 31--36.
\newline\urlprefix\url{https://ceur-ws.org/Vol-3597/paper6.pdf}

\bibitem{park2019specaugment}
D.~S. Park, W.~Chan, Y.~Zhang, C.-C. Chiu, B.~Zoph, E.~D. Cubuk, Q.~V. Le, Specaugment: A simple data augmentation method for automatic speech recognition, arXiv preprint arXiv:1904.08779 (2019).
\newblock \href {https://doi.org/10.48550/arXiv.1904.08779} {\path{doi:10.48550/arXiv.1904.08779}}.

\bibitem{fairseq}
M.~Ott, S.~Edunov, A.~Baevski, A.~Fan, S.~Gross, N.~Ng, D.~Grangier, M.~Auli, fairseq: A fast, extensible toolkit for sequence modeling, in: Proceedings of the 2019 Conference of the North American Chapter of the Association for Computational Linguistics (Demonstrations), 2019, pp. 48--53.
\newblock \href {https://doi.org/10.48550/arXiv.1904.01038} {\path{doi:10.48550/arXiv.1904.01038}}.

\bibitem{conneau2020unsupervised}
A.~Conneau, A.~Baevski, R.~Collobert, A.~Mohamed, M.~Auli, Unsupervised cross-lingual representation learning for speech recognition, arXiv preprint arXiv:2006.13979 (2020).
\newblock \href {https://doi.org/10.48550/arXiv.2006.13979} {\path{doi:10.48550/arXiv.2006.13979}}.

\bibitem{li2025survey}
M.~Li, Y.~Ahmadiadli, X.-P. Zhang, A survey on speech deepfake detection, ACM Computing Surveys 57~(7) (2025) 1--38.
\newblock \href {https://doi.org/10.1145/3714458} {\path{doi:10.1145/3714458}}.

\end{thebibliography}

\appendix
\section{Full results with minDCF Results}
\label{appendix:rawboost-minDCF}

While the official metric for the ASVSpoof5 Challenge is minDCF, our main paper reports EER due to its widespread use for deepfake speech detection. For completeness, we report minDCF in this appendix for multiple models, using the ASVSpoof5 cost configuration: $C_\mathrm{miss} = 1$, $C_\mathrm{fa} = 10$, and $\pi_\mathrm{spoof} = 0.05$.

\begin{table}[h!]
\centering
\scriptsize
\begin{tabular}{@{}llccc@{}}
\toprule
\textbf{Pretraining} & \textbf{Augmentations} & \textbf{EER Dev (\%)} & \textbf{EER Eval (\%)} & \textbf{minDCF Eval} \\
\midrule

\multirow{6}{*}{XLS-R 53} 
& $MS_B$ + C           & 1.36 & 4.53 & 0.127 \\
& $MS_B$ + R           & 2.29 & 4.82 & 0.130 \\
& $MS_B$ + C + R       & 2.86 & 4.99 & 0.139 \\
& $MF_G$ + C           & 1.38 & 5.22 & 0.148 \\
& $MF_G$ + R           & 3.26 & 5.21 & 0.146 \\
& $MF_G$ + C + R       & 1.88 & 4.94 & 0.138 \\

\midrule

\multirow{6}{*}{XLS-R 128} 
& $MS_B$ + C           & 0.34 & 4.75 & 0.137 \\
& $MS_B$ + R           & 1.82 & 4.17 & 0.119 \\
& $MS_B$ + C + R       & 2.39 & 4.22 & 0.121 \\
& $MF_G$ + C           & 0.29 & 5.01 & 0.144 \\
& $MF_G$ + R           & 1.97 & 4.86 & 0.139 \\
& $MF_G$ + C + R       & 1.65 & 4.31 & 0.121 \\

\midrule

\multirow{6}{*}{WAVLM} 
& $MS_B$ + C           & 0.86 & 4.77 & 0.132 \\
& $MS_B$ + R           & 3.64 & 5.54 & 0.156 \\
& $MS_B$ + C + R       & 1.69 & 5.40 & 0.154 \\
& $MF_G$ + C           & 2.57 & 6.96 & 0.188 \\
& $MF_G$ + R           & 3.62 & 5.51 & 0.153 \\
& $MF_G$ + C + R       & 2.44 & 5.38 &  0.154   \\

\midrule

\multirow{6}{*}{ASVSpoof5+}
& $MS_B$ + C           & 5.12 & 5.44 & 0.117 \\
& $MS_B$ + R           & 17.20 & 12.23 & 0.334 \\
& $MS_B$ + C + R       & 17.87 & 14.89 & 0.384 \\
& $MF_G$ + C           & 5.62 & 5.27 & 0.144 \\
& $MF_G$ + R           & 17.23 & 10.25 & 0.249 \\
& $MF_G$ + C + R       & 20.33 & 12.90 & 0.299 \\

\bottomrule
\end{tabular}
\caption{EER and minDCF results across different pretraining backbones and augmentations.}
\label{tab:minDCF}
\end{table}

In table \ref{tab:minDCF} we can see that EER and minDCF generally follow similar trends across models, indicating consistent relative performance. However, a notable exception emerges with ASVSpoof5+ trained with $MS_B$ + $C$, which achieves the lowest minDCF (0.117) but ranks outside the top ten in terms of EER (5.44\%). This discrepancy highlights how minDCF, influenced by decision thresholds and cost priors, can favor models with better-calibrated outputs over those with the lowest raw error rates.

\end{document}